\documentclass[journal]{IEEEtran}
\ifCLASSINFOpdf
  \usepackage[pdftex]{graphicx}
\else
\fi

\usepackage{amsmath}
\usepackage{array}

\begin{document}
%
\title{A Tight Upper Bound on Mutual Information}

\author{Michal~Hled\'ik,
        Thomas~R.~Sokolowski,
        and~Ga\v{s}per~Tka\v{c}ik

\thanks{The authors are with the Institute of Science and Technology Austria (IST Austria), Klosterneuburg
3400, Austria. E-mail: mhledik@ist.ac.at.}%
\thanks{This project has received funding from the European Union’s Horizon 2020 research and innovation programme under the Marie Sk\l{}odowska-Curie Grant Agreement No. 665385.}%
}

\maketitle

\begin{abstract}
We derive a tight lower bound on equivocation (conditional entropy), or equivalently a tight upper bound on mutual information between a signal variable and channel outputs. The bound is in terms of the joint distribution of the signals and maximum a posteriori decodes (most probable signals given channel output). As part of our derivation, we describe the key properties of the distribution of signals, channel outputs and decodes, that minimizes equivocation and maximizes mutual information. This work addresses a problem in data analysis, where mutual information between signals and decodes is sometimes used to lower bound the mutual information between signals and channel outputs. Our result provides a corresponding upper bound.
\end{abstract}


\section{Introduction}

\IEEEPARstart{T}{he} relationship between conditional entropy (equivocation) or mutual information, and best possible quality of decoding is an important concept in information theory. The best possible quality of a decoding scheme, when quantified by the minimal probability of error $\epsilon$, does not uniquely determine the value of equivocation or mutual information, but various upper and lower bounds have been proved, see Sec. \ref{sec_existing_theory}.

Here we discuss a scenario when not only $\epsilon$, but the complete joint probability distribution $p(x,\hat{x})$ of signals $x$ and maximum a posteriori 
decodes $\hat{x}$ is available. We refer to $p(x,\hat{x})$ as the confusion matrix. To our knowledge, such a scenario has not been extensively studied in the literature, despite having practical relevance for estimation of mutual information, as we point out in Sec. \ref{sec_motivation}. In this article, we derive an upper bound on mutual information (and a corresponding lower bound on equivocation) that is based on the confusion matrix and is tighter than the known similar bound by Kovalevsky and others \cite{Kovalevsky1968,Tebbe1968,Feder1994} based on probability of error alone. The inequality in our bound can be proved quickly using the bound by Kovalevsky, as we show in Sec. \ref{sec_quick_proof}. However, we also include a self-contained derivation in Sec. \ref{sec_our_proof}, where we construct the distribution of channel outputs that minimizes equivocation $H(X|Y)$ under our constraints.

 

\subsection{Equivocation, mutual information and the minimal probability of error}
\label{sec_existing_theory}

We consider a signal variable (message) $X$ that is communicated through a channel with output $Y$ and then decoded, obtaining a ``decode'' $\hat{X}$ -- forming a Markov chain $X \leftrightarrow Y \leftrightarrow \hat{X}$. The equivocation $H(X|Y)$ quantifies the uncertainty in $X$ if the value of $Y$ is given. Conversely, the mutual information $I(X;Y)$ measures how much information about $X$ is contained in $Y$. It is not surprising that both $H(X|Y)$ and $I(X;Y)$ can be related to the minimal probability of error while decoding, $\epsilon = \Pr(X \neq \hat{X})$.

Accurate decoding, i.e., low $\epsilon$, requires sufficiently low equivocation $H(X|Y)$. This is quantified by Fano's inequality \cite{Cover2006}. The mutual information between the true signal and the channel output,  $I(X;Y) = H(X)-H(X|Y)$, needs to be sufficiently high, and this is described by rate-distortion theory \cite{Shannon1959}.

\begin{figure}[!t]
\centering
\includegraphics[width=2.7in]{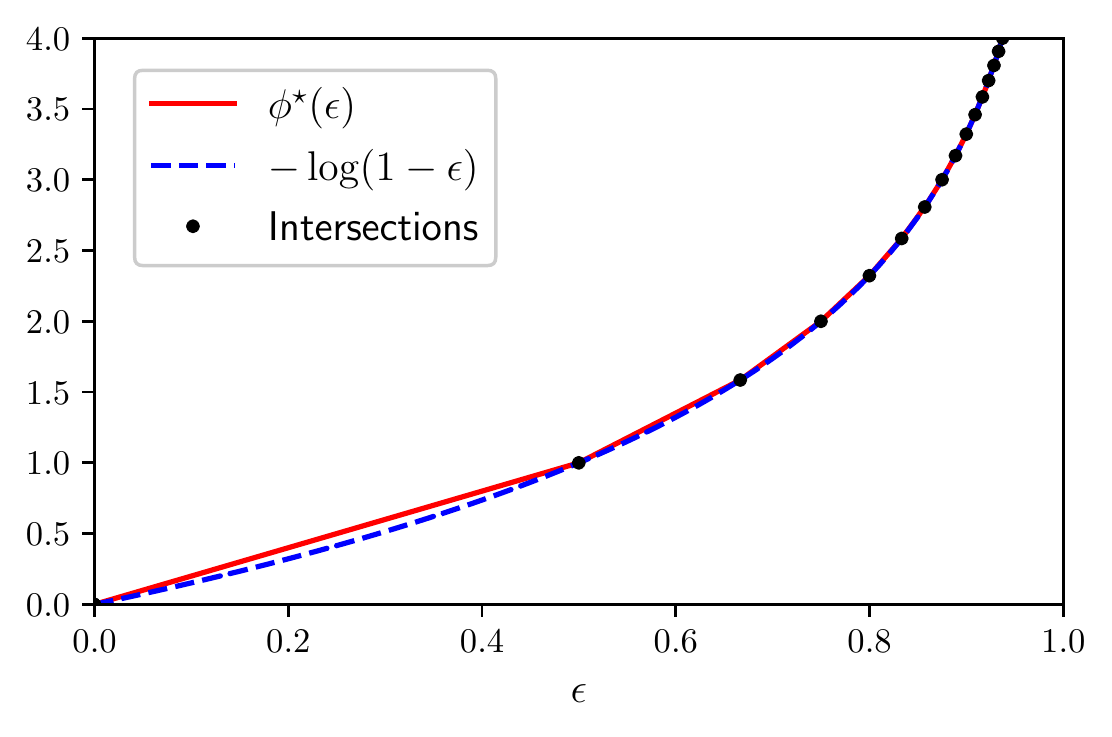}
\caption{Plot of the functions $\phi^\ast(\epsilon)$ and $-\log{(1-\epsilon)}$. The two functions intersect at $\epsilon=0$, $1/2$, $2/3$, $\dots$, $(|\mathcal{X}|-1)/|\mathcal{X}|$ (black dots), and in between $\phi^\ast(\epsilon)$ is piecewise linear.}
\label{fig_phi}
\end{figure}

Here we focus on the opposite bounds. If the minimal probability of error $\epsilon$ is specified, there is also a minimal possible equivocation. The following lower bound was derived for discrete $X$ with finite support by Kovalevsky \cite{Kovalevsky1968} and later Tebbe and Dwyer \cite{Tebbe1968} and Feder and Merhav \cite{Feder1994}. It reads
\begin{equation}
H(X|Y) \geq \phi^\ast (\epsilon), \label{eq_feder_merhav}
\end{equation}
where $\phi^\ast (\epsilon)$ is a piecewise linear function that coincides with $-\log{(1-\epsilon)}$ at points $\epsilon=0$, $1/2$, $2/3$, $\dots$, $(|\mathcal{X}|-1)/|\mathcal{X}|$ (we use $\log = \log_2$ throughout the paper, and $\mathcal{X}$ is the support of $X$), and it can be written using the floor and ceiling functions,
\begin{align}
\phi^\ast (\epsilon) &= \alpha(\epsilon) \log{ \left\lfloor \frac{1}{1- \epsilon} \right\rfloor} + \left(1-\alpha(\epsilon) \right) \log{ \left\lceil \frac{1}{1- \epsilon} \right\rceil}, \label{eq_phi} \\
\alpha(\epsilon) &= \left\lfloor \frac{1}{1- \epsilon} \right\rfloor \left( (1-\epsilon) \left\lceil \frac{1}{1- \epsilon} \right\rceil -1  \right). \label{eq_alpha}
\end{align}
The function $\phi^\ast (\epsilon)$ is plotted in Fig. \ref{fig_phi}.

The bound \eqref{eq_feder_merhav} has been generalized to countably infinite support of $X$ by Ho and Verd\'u \cite{Ho2010}. Sason and Verd\'u \cite{Sason2018} proved a generalisation of \eqref{eq_feder_merhav} for Arimoto-R\'enyi conditional entropy of arbitrary order. 

The bound \eqref{eq_feder_merhav} is tight when only the overall probability of error $\epsilon$ is available. However, when more constraints on the the joint distribution of $X$ and $Y$ are given, tighter bounds can be obtained. Prasad \cite{Prasad2015} introduced two series of lower bounds on $H(X|Y)$ based on partial knowledge of the posterior distribution $p(x|y)$. The first is in terms of the $k$ largest posterior probabilities $p(x|y)$ for each $y$, that we could label $p_1(y), p_2(y), \dots, p_k(y)$ in descending order (where $1 \leq k \leq |\mathcal{X}|$). The second series of bounds by Prasad is in terms of the averages of $p_1(y), p_2(y), \dots, p_k(y)$ across all $y$.

Hu and Xing \cite{Hu2016} focused on binary signal $X$ and derived a bound tighter than \eqref{eq_feder_merhav} by taking into account the prior distribution of signals $p(x)$. Hu and Xing also discuss suboptimal (other than maximum a posteriori) decoding, which is otherwise rare in the related literature.

\subsection{Motivation: estimation of mutual information} \label{sec_motivation}

Here we extend the bound \eqref{eq_feder_merhav} to account for the situation when the complete confusion matrix -- the joint distribution $p(x,\hat{x})$ is known. We are motivated by the following scenario: suppose that the goal is to estimate the mutual information $I(X;Y)$ from a finite set of $(x,y)$ samples. Moreover, assume that the space of possible channel outputs $\mathcal{Y}$ is large (much larger than the space of signals, $|\mathcal{Y}|\gg|\mathcal{X}|$), making a direct calculation of $I(X;Y)$ by means of their joint distribution $p(x,y)$ infeasible due to insufficient sampling. In such a case, one approach (used e.g. in neuroscience \cite{Borst1999}) is to construct a decoder, map each $y$ into a decode $\hat{x}$ and estimate the confusion matrix $p(x,\hat{x})$. Then the post-decoding mutual information $I(X;\hat{X})$ can be calculated and used as a lower bound on $I(X;Y)$ due to the data processing inequality \cite{Cover2006}. However, the gap between $I(X;\hat{X})$ and $I(X;Y)$ is not known (but see a discussion of this gap in \cite{Samengo2002}), and an upper bound on $I(X;Y)$ based on $p(x,\hat{x})$ is desirable. Our result is such a bound, for the specific case of maximum a posteriori decoder.

While mutual information $I(X;Y)$ has this practical importance, we formulate our result as an equivalent lower bound on equivocation $H(X|Y) = H(X)-I(X;Y)$ first. This is simpler to state and prove.

\section{Statement of the bound}

Given the joint distribution $p(X,\hat{X})$ of signals $X$ (discrete with finite support) and maximum a posteriori decodes $\hat{X}$ based on the channel output $Y$, the equivocation $H(X|Y)$ is bounded from below by
\begin{equation}
H(X|Y) \geq \sum_{\hat{x}} p(\hat{x}) \, \phi^\ast (\epsilon_{\hat{x}} ), \label{eq_bound_equivocation}
\end{equation}
where $\epsilon_{\hat{x}} = p(X \neq \hat{X} | \hat{x}) = 1 - p(X = \hat{x} | \hat{X} = \hat{x})$ is the probability of error for the decode $\hat{x}$ and the function $\phi^\ast$ is defined in \eqref{eq_phi}, \eqref{eq_alpha}.

Equivalently, we can bound the mutual information $I(X;Y)$ from above:
\begin{align}
I(X;Y) &= H(X) - H(X|Y)  \nonumber \\
 &\leq H(X) - \sum_{\hat{x}} p(\hat{x}) \, \phi^\ast (\epsilon_{\hat{x}} ).
\end{align}

These bounds are tight, and we construct the distributions $p(y|\hat{x})$ and $p(x|y)$ that achieve equality in Sec. \ref{sec_our_proof}.

\subsection{Comments on the bound}

We note that since the function $\phi^\ast (\epsilon_{\hat{x}})$ is convex, we can apply Jensen's inequality to the right hand side of \eqref{eq_bound_equivocation} and recover the bound \eqref{eq_feder_merhav} by Kovalevsky \cite{Kovalevsky1968},
\begin{equation}
H(X|Y) \geq \phi^\ast \left( \sum_{\hat{x}} p(\hat{x}) \, \epsilon_{\hat{x}} \right)
= \phi^\ast (\epsilon).
\end{equation}

Both bounds coincide in case of binary signal $|\mathcal{X}| = 2$, or any other case when the probability of error is less than $1/2$, $\epsilon_{\hat{x}} < 1/2$ for all $\hat{x}$. On this range, $\phi^\ast (\epsilon_{\hat{x}}) = 2 \epsilon_{\hat{x}}$ and the bound simplifies to
\begin{equation}
H(X|Y) \geq 2 \sum_{\hat{x}} p(\hat{x}) \, \epsilon_{\hat{x}} = 2 \epsilon,
\end{equation}
as has been noted in \cite{Feder1994} and before.

\subsection{Example calculation}

As an illustration, we apply our bound \eqref{eq_bound_equivocation} to an example confusion matrix and compare it to the bound \eqref{eq_feder_merhav} that is in terms of error probability $\epsilon$ only.

\begin{figure}[!t]
\centering
\begin{tabular}{m{1.3 in} m{1.8 in}}
    \includegraphics[width=\linewidth]{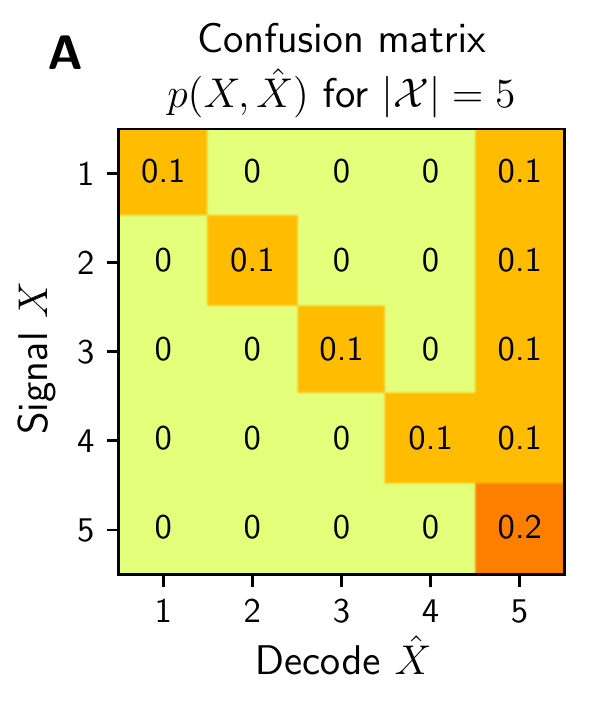} & 
    \includegraphics[width=\linewidth]{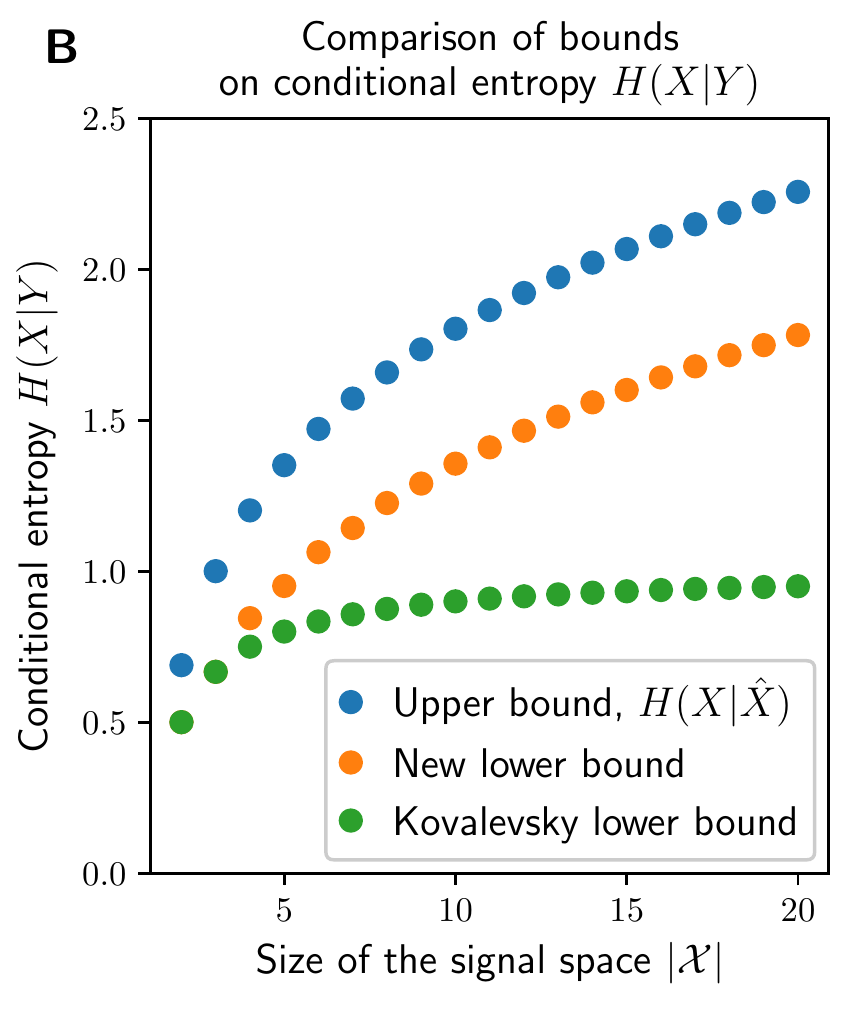}
\end{tabular}
\caption{Example application of the bound. (A) The joint distribution of signals and decodes $p(x,\hat{x})$ for which we compute the bound, defined in Eq. \eqref{eq_example_calculation}. Here for the case $|\mathcal{X}|=5$. (B) Bounds on conditional entropy (equivocation) $H(X|Y)$ plotted for different sizes of signal space $|\mathcal{X}|$. $H(X|Y)$ is bounded from above by $H(X|\hat{X})$ (blue points). Our novel lower bound (Eq. \eqref{eq_bound_equivocation}) is in orange and the bound by Kovalevsky (Eq. \eqref{eq_feder_merhav}) in green. Our bound \eqref{eq_bound_equivocation} is the tightest possible given the confusion matrix.}
\label{fig_simpleCalc}
\end{figure}

The confusion matrix considered is depicted in Fig. \ref{fig_simpleCalc} (A) for the case $|\mathcal{X}|=5$. We vary the size $|\mathcal{X}|$ of the space of signals $\mathcal{X}=\{ 1,2,\dots,|\mathcal{X}|\}$, and the confusion matrix always takes the form
\begin{equation}
    p(x,\hat{x}) = 
    \begin{cases}
    \frac{1}{2|\mathcal{X}|}; \qquad  &  x=\hat{x}<|\mathcal{X}|, \\
    \frac{1}{2|\mathcal{X}|}; \qquad  &  x<|\mathcal{X}|,\, \hat{x}=|\mathcal{X}|, \\
    \frac{1}{|\mathcal{X}|};  \qquad  &  x=\hat{x}=|\mathcal{X}|, \\
    0;                        \qquad  &  x \neq \hat{x},\, \hat{x}<|\mathcal{X}|.
  \end{cases} \label{eq_example_calculation}
\end{equation}
This distribution has the property that while most of the decodes have zero probability of being incorrect ($\epsilon_{\hat{x}}=0$ for $\hat{x}<|\mathcal{X}|$), the last one has a high probability of being incorrect, $\epsilon_{\hat{x}}=(|\mathcal{X}|-1)/(|\mathcal{X}|+1)$ for $\hat{x}=|\mathcal{X}|$. Our bound \eqref{eq_bound_equivocation} takes this into account -- which makes it substantially tighter than the bound \eqref{eq_feder_merhav} based only on the overall probability of error $\epsilon$. This can be seen in Fig. \ref{fig_simpleCalc} (B), where both lower bounds are plotted. We also plot the post-decoding conditional entropy $H(X|\hat{X})$ which serves as the upper bound on the true value of $H(X|Y)$.

\section{Proof of the bound}

We offer two alternative proofs of the bound here. The first proves it as a simple consequence of the bound \eqref{eq_feder_merhav} by Kovalevsky. It is short, but it leaves open the question of tightness. We therefore focus on the second proof, which is self-contained, implies tightness and perhaps offers additional insights, since it includes a derivation of the distribution of channel outputs $p(y|\hat{x})$, $p(x|y)$ that minimizes $H(X|Y)$.

Throughout the proofs, the spaces of possible values of $X$ and $Y$ are written as $\mathcal{X}$ and $\mathcal{Y}$ respectively. The decoding function is denoted $g : \mathcal{Y} \rightarrow \mathcal{X}$ and is based on the maximum a posteriori rule, $g(y) \in \underset{x}{\operatorname{argmax}} \, p(x|y)$. Finally, $\mathcal{Y}_{\hat{x}} = \{ y \in \mathcal{Y} \, | \, g(y) = \hat{x} \}$ is the set of all $y$ that decode into $\hat{x}$.

\subsection{A quick proof of inequality following Kovalevsky's bound} \label{sec_quick_proof}

The left hand side of \eqref{eq_bound_equivocation}, the equivocation $H(X|Y)$ can be written as
\begin{equation}
H(X|Y) = \sum_{\hat{x}} p(\hat{x}) \int_{\mathcal{Y}_{\hat{x}}} H(X|Y=y) \,dp(y|\hat{x}),
\end{equation}
where the term $\int_{\mathcal{Y}_{\hat{x}}} H(X|Y=y) \,dp(y|\hat{x})$ is the entropy of $X$ conditional on $Y$, but with the values of $Y$ only limited to $\mathcal{Y}_{\hat{x}}$. Since it has the form of conditional entropy, we can use the Kovalevsky bound \eqref{eq_feder_merhav} and obtain our result \eqref{eq_bound_equivocation}.

This establishes the inequality in our bound, but it does not tell us if equality can be achieved -- and if it can, for what distribution of $Y$ does it happen. We address this in the following section.

\section{Proof by minimization of equivocation}
\label{sec_our_proof}

For simplicity, we formulate the derivation for discrete $Y$. However, as we comment in Sec. \ref{sec_discussion}, the derivation applies to continuous $Y$ with only minor modifications.

For clarity, let us state the minimization problem we are solving. We minimize 
\begin{align}
H(X|Y) = \sum_{\hat{x}} p(\hat{x}) \sum_{y \in \mathcal{Y}_{\hat{x}}} p(y|\hat{x}) H(X|Y=y) \label{eq_objective_function_full}
\end{align}
with respect to $p(y|\hat{x})$ and $p(x|y)$, with the constraints given by the confusion matrix and maximum a posteriori decoding:
\begin{align}
\forall x, \hat{x}: \qquad & \sum_y p(x|y) p(y|\hat{x}) = p(x|\hat{x}), \label{eq_constraints_conf} \\
\forall \hat{x}, \forall y \in \mathcal{Y}_{\hat{x}}: \qquad & \hat{x} \in \underset{x}{\operatorname{argmax}} \, p(x|y). \label{eq_constraints_dec}
\end{align}

Note in \eqref{eq_objective_function_full} that the minimization can be done separately for each $\hat{x}$, since the corresponding $\mathcal{Y}_{\hat{x}}$ are disjoint. Hence we have $|\mathcal{X}|$ independent minimization problems with the objective function
\begin{equation}
\sum_{\mathcal{Y}_{\hat{x}}} p(y|\hat{x}) H(X|Y=y). \label{eq_objective_function}
\end{equation}

Note also that we do not have any constraint on $|\mathcal{Y}|$, the number of elements of $\mathcal{Y}$. We actually exploit this flexibility in the proof. However, it turns out (see Propositions 1 and 2) that when the minimum is achieved, there can be only a limited number of $y$ values with different distribution $p(x|y)$.

Our approach is based on update rules for $p(y|\hat{x})$ and $p(x|y)$ that decrease the objective function \eqref{eq_objective_function} while respecting the constraints \eqref{eq_constraints_conf}, \eqref{eq_constraints_dec}. In fact, the updates also change $|Y|$. The minimum of $H(X|Y)$ is achieved when the update rules can no longer be used to decrease it -- and such situations can be characterized and the corresponding $H(X|Y)$ can be calculated.

It is instructive to have in mind the following visualization of our minimization problem, which we use to illustrate the update rules in Fig. \ref{fig_updateRules}. The distribution $p(x,y|\hat{x})$ for some $\hat{x}$, with $y$ restricted to $y \in \mathcal{Y}_{\hat{x}}$ can be represented as a matrix, with a row for each $x$ and a column for each $y$. Normalized columns correspond to $p(x|y)$ and the sum of each column is $p(y|\hat{x})$. The constraint \eqref{eq_constraints_conf} means that each row has a fixed sum, $p(x|\hat{x})$, and the constraint \eqref{eq_constraints_dec} means that one row (e.g. the first) contains the dominant elements of all columns. The objective function \eqref{eq_objective_function} is a weighted sum of entropies of all columns. Our minimization will consist of adding and removing columns, and moving probability mass within rows.

In the following, a probability distribution is called \emph{flat} if all non-zero elements are equal, i.e. there are $n$ non-zero elements and all have probabilities $1/n$. The number $n$ is called its \emph{length}.

\subsection*{Proposition 1: equivocation minimized by flat $p(x|y)$}
\label{subsec_proposition1}

The minimum of the objective function \eqref{eq_objective_function}, given constraints \eqref{eq_constraints_conf}, \eqref{eq_constraints_dec} can only be achieved when the distributions $p(x|y)$ are flat for all $y$.

\begin{figure}[!t]
\centering
\includegraphics[width=2.8in]{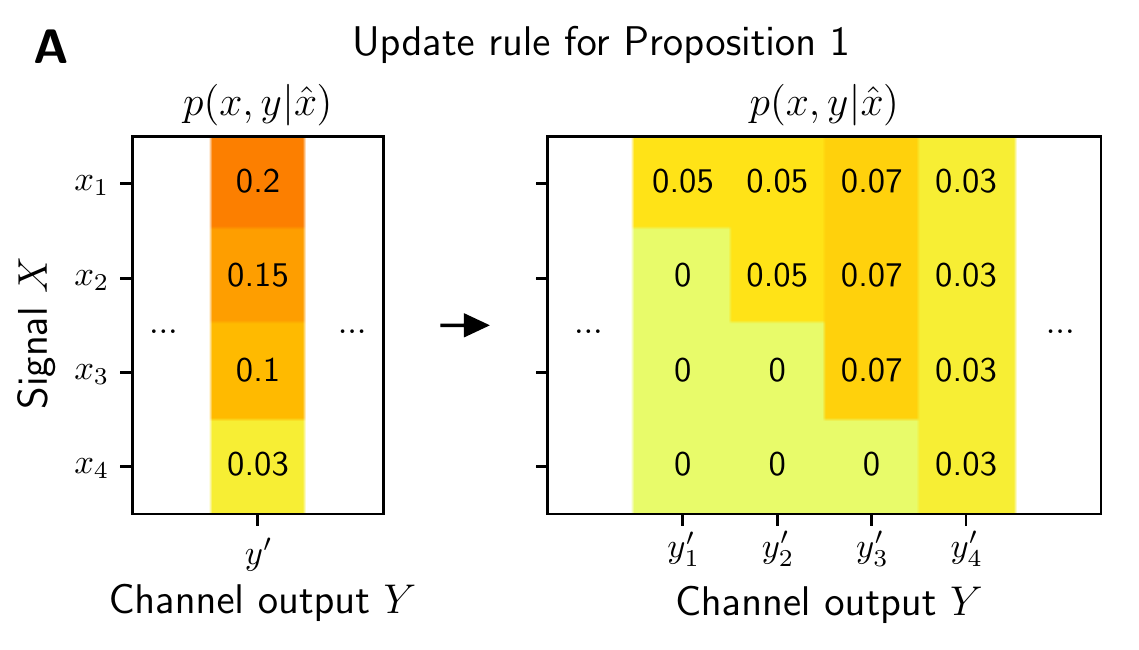}\\
\includegraphics[width=2.8in]{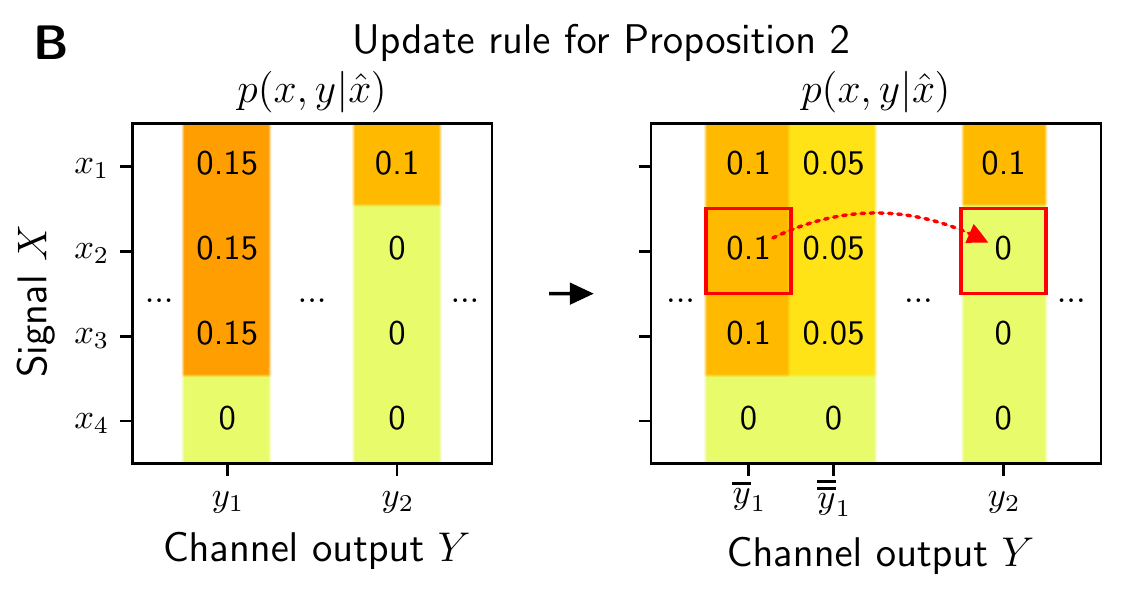}
\caption{Illustrations of the update rules used to prove (A) Proposition 1 and (B) Proposition 2. Displayed is the joint distribution $p(x,y|\hat{x})$. 
(A) A channel output $y'$ with a non-flat distribution $p(x|y')$ is replaced by $y'_1, y'_2, \dots, y'_{4}$ with flat distributions $p(x|y'_i)$, such that $y'_1, y'_2, \dots, y'_{4}$ still decode into $x_1$ and the confusion matrix is not affected. This replacement decreases $H(X|Y)$, our objective function. The elements of $\mathcal{X}$ are labeled in decreasing order of $p(x,y|\hat{x})$.
(B) Two channel outputs, $y_1$ and $y_2$, have flat distributions $p(x|y_{1,2})$ with $3$ and $1$ nonzero elements respectively. We replace $y_1$ by $\bar{y}_1$ and $\bar{\bar{y}}_1$, and then transfer probability $p(x_2,\bar{y}_1|\hat{x})$ to $p(x_2,y_2|\hat{x})$ (dotted red arrow). The distributions $p(x|\bar{y}_1)$, $p(x|\bar{\bar{y}}_1)$ and $p(x|y_{2})$ remain flat, and the objective function $H(X|Y)$ is decreased.
} 
\label{fig_updateRules}
\end{figure}

\begin{IEEEproof}
Suppose that there is a channel output $y'$ with a non-flat distribution $p(x|y')$. Then, the following update rule, illustrated in Fig. \ref{fig_updateRules} (A), will decrease the objective function \eqref{eq_objective_function}.

We label the elements of $\mathcal{X}$ as $x_1, x_2, \dots, x_{|\mathcal{X}|}$ such that
\begin{equation}
p(x_1|y') \geq p(x_2|y') \geq \dots \geq p(x_{|\mathcal{X}|}|y') \geq 0,
\end{equation}
where at least two of the inequalities are sharp (otherwise $p(x|y')$ would be flat). Note that $x_1$ must be the decode of $y'$, i.e. $g(y') = \hat{x} = x_1$. The proposed update is to replace $y'$ by $y'_1, y'_2, \dots, y'_{|\mathcal{X}|}$ with flat distributions $p(x|y'_i)$,
\begin{align}
p(x_j|y'_i) &= \begin{cases}
1/i; & j \leq i \\
0; & j > i,
\end{cases} \label{eq_prop1_xy} \\
p(y'_i|\hat{x}) &= \begin{cases} 
i p(y'|\hat{x})  \left( p(x_i|y') - p(x_{i+1}|y') \right); & i < |\mathcal{X}| \\
i p(y'|\hat{x}) \, p(x_i|y'); & i = |\mathcal{X}|.
\end{cases}  \label{eq_prop1_yx}
\end{align}
Intuitively, this replaces $y'$ by multiple elements $y'_i$ with flat distributions $p(x|y'_i)$ covering the first $1, 2, \dots, |\mathcal{X}|$ elements of the ordered $x_1, x_2, \dots, x_{|\mathcal{X}|}$. It can be confirmed that this replacement respects the constraints \eqref{eq_constraints_conf}. All $y'_i$ still decode into $\hat{x} = x_1$, and the probability associated with $y'$ is merely divided among the elements $y'_i$,
\begin{align}
\sum_i p(y'_i|\hat{x}) &= p(y'|\hat{x}), \label{eq_prop1_argument1} \\
\sum_i p(x_j|y'_i) p(y'_i|\hat{x}) &= p(x_j|y') p(y'|\hat{x}). \label{eq_prop1_argument2}
\end{align}
See Fig. \ref{fig_updateRules} for an example.

Now we look at the change in the objective function \eqref{eq_objective_function} induced by this replacement. Before the replacement, $y'$ contributes the amount
\begin{equation}
p(y'|\hat{x}) H(X|Y=y'), \label{eq_prop1_oldH}
\end{equation}
where $H(X|Y=y')$ is the entropy of a single random variable with distribution $p(x|y')$. After the replacement, the total contribution of all $y'_1, y'_2, \dots, y'_{|\mathcal{X}|}$ is
\begin{multline}
\sum_i p(y'_i|\hat{x}) H(X|Y=y'_i) = \\
= p(y'|\hat{x}) \sum_i \frac{p(y'_i|\hat{x})}{p(y'|\hat{x})} H(X|Y=y'_i), \label{eq_prop1_newH}
\end{multline}
where the latter sum has the form of a conditional entropy of a variable with a marginal distribution $p(x|y')$, conditioned on the value of $y'_i$ distributed according to $p(y'_i|\hat{x})/p(y'|\hat{x})$ (this follows from Eq. \eqref{eq_prop1_argument1}, \eqref{eq_prop1_argument2}). Since conditioning decreases entropy, our replacement decreases the objective function \eqref{eq_objective_function}.

The only case when our the proposed replacement cannot be used to decrease the objective function is when $p(x|y)$ is flat for all $y$. Therefore flat $p(x|y)$ must be a characteristic of any solution to our minimization problem.
\end{IEEEproof}

Note that there are only $2^{|\mathcal{X}|-1}$ different possible flat distributions $p(x|y)$ with nonzero $p(X=\hat{x}|y)$, which means that we need at most $2^{|\mathcal{X}|-1}$ elements in $\mathcal{Y}_{\hat{x}}$ to achieve the minimum equivocation. However, as the following proposition will show, there are further restrictions on $p(x|y)$ at the minimum.

Reflecting that only flat $p(x|y)$ are of further interest in the minimization, we say that the channel output $y$ has length $l$ if $p(x|y)$ has length $l$.

\subsection*{Proposition 2: minimization restricts lengths of $p(x|y)$ }
\label{subsec_proposition2}

Building on Proposition 1, we further claim that if equivocation is minimized, no two channel outputs $y_1,y_2 \in \mathcal{Y}_{\hat{x}}$ can have lengths differing by more than $1$.

\begin{IEEEproof}
As before, we introduce an update rule. Recalling the visualization with a column for each $y$, this update rule will move a nonzero element from a longer column to a shorter column, as shown in Fig. \ref{fig_updateRules} (B).

Take two elements $y_1,y_2 \in \mathcal{Y}_{\hat{x}}$ that have flat distributions $p(x|y_1)$ and $p(x|y_2)$ with lengths $a$ and $b$ respectively where $a>b$. Assume that $a$ and $b$ differ by more than one, $a - b > 1$. This means that we can choose an element $x' \in \mathcal{X}$ such that $p(x'|y_1) = 1/a$ and $p(x'|y_2) = 0$. Assume momentarily that $p(y_1|\hat{x})/a = p(y_2|\hat{x})/b$ (we will relax this assumption later). Then we can replace $y_1,y_2$ by $y'_1$ and $y'_2$, such that
\begin{itemize}
\item $p(x|y'_1)$ is flat with length $a-1$. It is nonzero for the same $x$ as $p(x|y_1)$, except for $x'$ where it is zero.
\item $p(x|y'_2)$ is flat with length $b+1$. It is nonzero for the same $x$ as $p(x|y_2)$, and also for $x'$.
\end{itemize}
Given that $p(y_1|\hat{x})/a = p(y_2|\hat{x})/b$, we can also choose the probabilities $p(y'_1|\hat{x})$ and $p(y'_2|\hat{x})$ such that $y'_1$, $y'_2$ contribute the same amount to $p(x|\hat{x}) = \sum_{y} p(x|y) p(y|\hat{x})$ as $y_1$ and $y_2$ did, ensuring that constraints \eqref{eq_constraints_conf} are respected:
\begin{align}
p(y'_1|\hat{x}) &= \frac{a-1}{a} p(y_1|\hat{x}), \\
p(y'_2|\hat{x}) &= \frac{b+1}{b} p(y_2|\hat{x}).
\end{align}

Now we show that the proposed replacement reduces the objective function. Before the replacement, the contribution to the objective function \eqref{eq_objective_function} by $y_1$ and $y_2$ was
\begin{equation}
p(y_1|\hat{x}) \log{a} + p(y_2|\hat{x}) \log{b}.
\end{equation}
After the replacement, $y'_1$ and $y'_2$ contribute by
\begin{equation}
\frac{a-1}{a} p(y_1|\hat{x}) \log{(a-1)} + \frac{b+1}{b} p(y_2|\hat{x}) \log{(b+1)}.
\end{equation}
The difference of these contributions $\Delta$, after minus before replacement, has the form
\begin{equation}
\Delta = \frac{p(y_1|\hat{x})} {a} \left(   f(b+1) - f(a) \right),
\end{equation}
where $f(t) = t \log{t} - (t-1) \log{(t-1)}$ is an increasing function for $t\geq 1$. Since $b+1 < a$, we have $\Delta < 0$, meaning that the objective function is reduced.

This update rule is applicable to any $y_1,y_2 \in \mathcal{Y}_{\hat{x}}$ with lengths $a$ and $b$ such that $a-b>1$ respectively. We have, however, further required that $p(y_1|\hat{x})/a = p(y_2|\hat{x})/b$. This requirement can be avoided. If $p(y_1|\hat{x})/a > p(y_2|\hat{x})/b$, we first split $y_1$ into $\bar{y}_1$ and $\bar{\bar{y}}_1$ with
\begin{align}
p(\bar{y}_1|\hat{x}) &= a \, p(y_2|\hat{x})/b, \\
p(\bar{\bar{y}}_1|\hat{x}) &= p(y_1|\hat{x}) - a \, p(y_2|\hat{x})/b, \\
p(x|\bar{y}_1) &= p(x|\bar{\bar{y}}_1) = p(x|y_1),
\end{align}
such that the above mentioned update rule can be applied to $\bar{y}_1$ and $y_2$ while $\bar{\bar{y}}_1$ is left unchanged, see Fig. \ref{fig_updateRules} (B). If $p(y_1|\hat{x})/a < p(y_2|\hat{x})/b$, we can proceed analogously by splitting $y_2$.

We can decrease the objective function by repeatedly applying this generalized update rule. Therefore, the minimum can only be achieved when the lengths of $p(x|y)$ for $y \in \mathcal{Y}_{\hat{x}}$ vary by no more than 1.
\end{IEEEproof}

Note that by repeated application of this update rule, in a finite number of steps we reach a state with only up to two lengths (per $\hat{x}$) that differ by at most 1. As shown in the next section, such a state implies a specific value of $H(X|Y)$. Together with the update rule in the proof of Proposition 1, this gives us an algorithm to find the distributions $p(y|\hat{x})$ and $p(x|y)$ that achieves the minimum $H(X|Y)$. The algorithm can start from an arbitrary initialization of $p(y|\hat{x})$ and $p(x|y)$ that follows the constraints \eqref{eq_constraints_conf}, \eqref{eq_constraints_dec} and finishes in a finite number of steps.


It remains to be determined what are the (at most two) allowed lengths of $y \in \mathcal{Y}_{\hat{x}}$ and how the elements $y$ with these lengths contribute to the equivocation $H(X|Y)$.

\subsection*{Admissible lengths of $p(x|y)$}

Let us call the two admissible lengths $l_{\hat{x}}$ and $l_{\hat{x}}+1$. Given $\hat{x}$, the total probability of all $y \in \mathcal{Y}_{\hat{x}}$ with length $l_{\hat{x}}$ is $\alpha_{\hat{x}}$, and those of length $l_{\hat{x}}+1$ have probability $1-\alpha_{\hat{x}}$. Then from the constraint \eqref{eq_constraints_conf}, we can write the probability that $\hat{x}$ is the correct decode 
\begin{equation}
1-\epsilon_{\hat{x}} = \frac{\alpha_{\hat{x}}}{ l_{\hat{x}} }+ \frac{1-\alpha_{\hat{x}}} {l_{\hat{x}}+1}, \label{eq_alpha_length}
\end{equation}
from which we can deduce that $\frac{1} {l_{\hat{x}}+1} \leq 1-\epsilon_{\hat{x}} \leq \frac{1}{ l_{\hat{x}} }$, and that the two admissible lengths must be 
\begin{equation}
l_{\hat{x}} = \left\lfloor \frac{1}{1- \epsilon_{\hat{x}}} \right\rfloor 
\text{ and } l_{\hat{x}} + 1 = \left\lceil \frac{1}{1- \epsilon_{\hat{x}}} \right\rceil, \label{eq_result_lengths}
\end{equation}
unless $\frac{1}{1- \epsilon_{\hat{x}}}$ is an integer -- in that case the floor and ceiling coincide into a single admissible length.

Now, from equations \eqref{eq_alpha_length} and \eqref{eq_result_lengths} we can determine that 
\begin{equation}
\alpha_{\hat{x}} = \left\lfloor \frac{1}{1- \epsilon_{\hat{x}}} \right\rfloor \left( (1-\epsilon_{\hat{x}}) \left\lceil \frac{1}{1- \epsilon_{\hat{x}}} \right\rceil -1  \right) = \alpha(\epsilon_{\hat{x}}) \label{eq_result_alpha}
\end{equation}
is the total probability (given $\hat{x}$) of $y \in \mathcal{Y}_{\hat{x}}$ with length $\lfloor \frac{1}{1- \epsilon_{\hat{x}}} \rfloor$.

Finally, the minimal value of equivocation is simply
\begin{equation}
H(X|Y) \geq \sum_{\hat{x}} p(\hat{x}) \left( \alpha_{\hat{x}} \log{l_{\hat{x}}} + (1-\alpha_{\hat{x}}) \log{(l_{\hat{x}} + 1)} \right),
\end{equation}
which together with equations \eqref{eq_result_lengths} and \eqref{eq_result_alpha} constitutes our main bound, as stated in \eqref{eq_bound_equivocation}.

\section{Discussion} \label{sec_discussion}

We have introduced a tight lower bound on equivocation in terms of the maximum a posteriori confusion matrix, and proved it in two ways. The first is a proof of the inequality, starting from a similar bound by Kovalevsky \cite{Kovalevsky1968}, but it does not prove that the bound is tight. Therefore, we developed a second proof, in which we construct the distribution of channel outputs that minimizes the equivocation and achieves equality in our bound.

Central to the latter approach are two update rules for the distribution of the channel outputs. These update rules exploit the fact that equivocation can be, under our constraints, minimized by (1) making the posterior distributions $p(x|y)$ flat and (2) making sure that these flat distributions contain similar numbers of nonzero elements.

We formulated the proof for discrete random variables $X$ and $Y$, but it can be extended. If $X$ is discrete but $Y$ continuous, application of a modified version of the first update rule would result in $2^{|\mathcal{X}|}$ regions in the $\mathcal{Y}_{\hat{x}}$ space corresponding to each of the $2^{|\mathcal{X}|}$ possible flat distributions $p(x|y')$. For example, the region associated with a flat distribution of length $|\mathcal{X}|$, that is $p(x|y') = 1/|\mathcal{X}|$, would have a total probability $\int_{\mathcal{Y}_{\hat{x}}} |\mathcal{X}| \min_x{p(x|y)} dp(y|\hat{x})$. These subsets of $\mathcal{Y}$ where $p(x|y)$ is constant can then be treated like discrete values, and the rest of our derivation applies.

Bounds on equivocation (or mutual information) in terms of the confusion matrix are, to our knowledge, not common -- despite their relevance for estimation of mutual information. We hope that our result can be useful for these purposes, and that it sheds some light on the gap between mutual information before and after decoding. However, its applicability is restricted by the assumption of maximum a posteriori decoding, and relaxing this assumption remains an interesting challenge.

\section*{Acknowledgment}

The authors would like to thank Sarah A. Cepeda-Humerez for helpful discussions.


%

\end{document}